\documentclass[amsmath,twocolumn,superscriptaddress,prl,aps]{revtex4}
\usepackage{lmodern}
\usepackage{lmodern}
\usepackage[latin9]{inputenc}
\usepackage{textcomp}
\usepackage{amsmath}
\usepackage{amssymb}
\usepackage{graphicx}
\usepackage{esint}
\usepackage{bm}
\usepackage{color}
\usepackage{epstopdf}
\makeatletter
\@ifundefined{textcolor}{}
{%
 \definecolor{BLACK}{gray}{0}
 \definecolor{WHITE}{gray}{1}
 \definecolor{RED}{rgb}{1,0,0}
 \definecolor{GREEN}{rgb}{0,1,0}
 \definecolor{BLUE}{rgb}{0,0,1}
 \definecolor{CYAN}{cmyk}{1,0,0,0}
 \definecolor{MAGENTA}{cmyk}{0,1,0,0}
 \definecolor{YELLOW}{cmyk}{0,0,1,0}
}
\newcommand{\beq} {\begin{equation}}
\newcommand{\eeq} {\end{equation}}
\newcommand{\bea} {\begin{eqnarray}}
\newcommand{\eea} {\end{eqnarray}}
\newcommand{\be} {\begin{equation}}
\newcommand{\ee} {\end{equation}}

\makeatother
\begin{document}

\title{Itinerant scenario for Fe-pnictides: comparison with quantum Monte Carlo}

\author{Andrey V. Chubukov and Rui-Qi Xing}
\affiliation{School of Physics and Astronomy, University of Minnesota,
Minneapolis, MN 55455, USA}
 \begin{abstract}

Recent applications of Quantum Monte Carlo (QMC) technique to Fe-based superconductors opened a way to directly verify the applicability of the itinerant scenario for these systems.  Fe-based superconductors undergo various instabilities upon lowering temperature (magnetism, superconductivity, nematicity/orbital order), and one can check whether  the hierarchy of instabilities obtained within the itinerant approach is the same as in unbiased QMC simulations.  In a recent paper [arXiv:1512:08523] the authors considered the simplest two-band model with interaction tailored to favor orbital order. The type of the orbital order found in QMC is different from the one found in earlier itinerant analysis.  We report the results of our calculations within the itinerant scenario and argue that they are in perfect agreement with QMC.
 \end{abstract}
\maketitle

\newpage

\section{Introduction.}

 The issue whether Fe-based  (FeSCs) can be viewed as fully itinerant electronic systems, or electrons from
 some of the orbitals  are localized, has been at the center of the debates on FeSCs right from their discovery\cite{mazin,kuroki,kemper,chubukov,fernandes_12,fs,fcs,kontani,phillips,kruger,ku,si,hu,medici,leni,gabi}.  The itinerant scenario is justified when the interactions are  smaller than the fermionic bandwidth, and it treats various instabilities in FeSCs, such as magnetism, superconductivity (SC), Ising-nematic spin order and spontaneous orbital order, as low energy instabilities are determined by carriers located near hole and/or electron Fermi surfaces (FSs) (Refs. \cite{kemper,chubukov,fernandes_12,fs,fcs,kontani}). Localized scenario, on the other hand, is  justified when the density-density (Hubbard) interactions are larger than the bandwidth, and within this scenario instabilities in FeSCs involve carriers from everywhere in the Brillouin zone~\cite{phillips,kruger,ku,si,hu,medici,leni}.
   The third, Hund metal scenario,
  has also been put forward~\cite{gabi} --  it assumes that Hund interaction is large enough, in which case the system retains a metallic behavior but becomes a bad metal.

  The FeSCs are metals for all dopings and compositions, and this would generally place them in the category of itinerant systems. At the same time the
    values of the Hubbard and Hund interactions in FeSCs, obtained from first-principle calculations,  are comparable to the  bandwidth,
     and some measurements of magnetic excitations in parent compounds have been reasonably well reproduced in calculations based on both itinerant~\cite{eremin} and localized~\cite{dai} scenario.
   Measurements of the specific heat in strongly hole-doped FeSCs (specifically in K$_{x}$ Ba$_{1-x}$ Fe$_2$ As${_2}$ for $x \to 1$, Refs. \cite{c_inkfeas})
   have been interpreted both within the itinerant scenario~\cite{efremov_2}, and by assuming that electrons on some of the Fe-orbitals get localized~\cite{medici,leni}, and mid-infrared optical data have been interpreted within the  Hund metal scenario~\cite{gabi_2}.
   Furthermore,  weak coupling and strong coupling scenarios for FeSCs yield  the same set of ordered states -- magnetism, SC, etc.  This all makes it difficult to settle on the approach. It also raises the fundamental issue whether FeSCs can be viewed as the "systems in the intermediate regime", which can be gradually reached starting from weak coupling, but some of the features they display are inherently  strong coupling and are completely missed in weak coupling calculations.

   To the best of our knowledge, there are no experimental data for weakly/moderately doped Fe-pnictides, which could not be reproduced, at least qualitatively, within the weak coupling scenario.
    In Fe-selenades, the situation is a bit more involved as magnetism in FeTe -- the parent compound of FeTe$_{1-x}$Se$_x$ family, is qualitatively different from that in other parent compounds of FeSCs  (double stripe or plaquette~\cite{natasha} in FeTe vs single stripe in other systems~\cite{stripe}) and is only reproduced within the localized scenario~\cite{natasha}.  Still, an orbital order at $x \approx 1$ and  SC at $x \geq 0.5$ in this material and also in 245 family of FeSCs are all reproduced within the itinerant approach~\cite{kontani,cfs_15,ckf,i_paul,khodas}.  This is in line with the idea that FeSCs, with the exception of FeTe, can be viewed as  itinerant systems, adiabatically extended to intermediate coupling.

However,  the ability to qualitatively explain the data may be misleading and one needs another tool to verify whether the behavior of a given FeSC, particularly the hierarchy of the instabilities upon lowering temperature or changing parameters, differs in any fundamental way from that in a weakly coupled metal with the same topology of the FS as in a FeSC. Recent application of Quantum Monte Carlo (QMC) method to FeSCs (Refs.~\cite{singh,ashvin}) provide such a tool as they allow one to compare the actual behavior in a fermionic system with comparable strength of kinetic  and potential energies with the one obtained by using weak coupling perturbative schemes  for various  topologies of Fermi surfaces  and structures of multi-band excitations.   In Ref.~\cite{ashvin}, Dmitriesku et al applied QMC to the simplest  toy model for FeSCs -- the two band model with Fe $d_{xz}$  and $d_{yz}$  orbitals, first considered by Raghu et al~\cite{raghu}.  Although this model does not reproduce the correct topology of low-energy electronic states in FeSCs  (one of the two hole FSs is in wrong place in the 1 Fe Brillouin zone), it nevertheless is the simplest toy model with two hole and two electron Fermi surfaces.  Dmitriesku et al further tailored 4-fermion interaction to be local and in the form  $-g (n_{xz} - n_{yz})^2$, where $n_i = d^\dagger_i d_i$
 is fermionic density on a given orbital, and $g >0$. This last condition is difficult to justify on microscopic grounds because it implies that intra-orbital Hubbard interaction is attractive, in variance with first-principle calculations~\cite{first_principles}.
   Nevertheless, the model considered in \cite{ashvin} and in earlier works~\cite{yamaze}  is quite interesting for our purpose to compare QMC and perturbation theory as it favors an orbital order (a spontaneous development of a non-zero $<n_{xz} -n_{yz}>$).  An orbital order is a threshold phenomenon, i.e., it  appears
 only when $g$ exceeds a certain critical value $g_c$, which is generally of order of the bandwidth, W.
 Weak coupling analysis is an expansion in $g$, and there is no a'priori guarantee that the type of orbital order obtained by using weak coupling  approximation and extending $g$ to critical $g_c$ will be the right one.

Different types  of  orbital order in the two-pocket model include ferro-orbital (FO) order  (a Pomeranchuk instability), which can be in $s-$wave or $d-$wave channel,
antiferro-orbital (AFO) order  with momentum $(\pi,\pi)$, and stripe-type orbital order with momentum $(0,\pi)$ or $(\pi,0)$.
 Dmitriesku et al argued that strong coupling
analysis (an expansion in $1/g$) favors AFO order, and they did find the same type of order in QMC.  Earlier itinerant calculations, on the other hand
 found numerically a different, FO order (Ref. \cite{yamaze}).

 \subsection{Summary of our results}

   In this paper we report the results of our analytical  analysis of the orbital order and SC  within the itinerant scenario.

  In the particle-hole channel we found  that weak coupling calculation shows that the system chooses to develop AFO order,  the same one as found in QMC calculations.  Moreover, we argue that some of the particle-hole polarization bubbles involved in the renormalizations of  the orbital order parameters are logarithmical at intermediate energies and, as a result, the critical coupling $g = g_c$ for the orbital instability is small compared to fermionic bandwidth, $W$,
  and is within the range of applicability of weak coupling expansion. The  smallness of $g_c/W$ is due to small sizes of hole and electron pockets and holds in $1/\log{\frac{W}{\epsilon_0}}$, where $\epsilon_0$ is of order of Fermi energy.
     We found  logarithms in both AFO and FO channels,
    but the prefactor in the AFO channel is larger, hence the leading instability is towards AFO  order.

In the particle-particle channel we found the leading instability in $s^{++}$ channel  (ordinary $s-$wave) and subleading instability in the $d-$wave channel.  The renormalizations in both channels contain series of conventional Cooper-type  $\frac{g}{W} \log{\frac{W}{T}}$ terms, but also contain terms of order $g/\epsilon_0$,
due to the presence of two weakly dispersing bands, whose energies remain of order $\epsilon_0$ over a wide range of momenta. S-wave channel wins over $d-$wave both at truly weak coupling, when $T_c \ll \epsilon_0$
 and only  the conventional logarithmical terms matter, while at larger $g$, terms of order $g/\epsilon_0$ play the leading role.

We compared critical $g$ for the instabilities in the particle-hole and particle-particle channels.
  Within the ladder approximation, when particle-particle and particle-hole channels do not couple to each other, the comparison of the eigenvalues in the AFO and $s^{++}$ channels shows that the overall prefactor in the AFO channel is  larger than in $s^{++}$ channel, but the combination of polarization operators is larger in the SC channel. We went beyond the ladder approximation and used renormalization group (RG) to include the flow of the intra-orbital and inter-orbital interactions
        between high and low  energies due to the actual presence of the couplings between particle-hole and particle-particle channels.
         We found that, due to the flow, the  overall prefactor in the AFO channel is reduced and becomes the same as in $s^{++}$ channel.
         Because polarization operator is larger in the SC channel,  the leading instability upon, e,g., increasing $g$ at a certain non-zero temperature is definitely towards   $s^{++}$ SC. The AFO order develops, but at a larger $g$.  This fully agrees with QMC calculations.

Another result of RG is that couplings in both $s^{++}$  and AFO channels get enhanced by coupling to stripe magnetic fluctuations. This enhancement is the strongest around half-filling, when there is nesting between hole and electron pockets~\cite{chubukov}. Accordingly, both  $s^{++}$ SC and AFO order are the strongest near half-filling. This again agrees with QMC results.

The structure of the paper is the following. In the next section we introduce the model. In Sec. \ref{sec:inst} we consider instabilities in the particle-hole and particle-particle channels and the interplay between them.
In Sec. \ref{sec:oo} we analyze instabilities towards  FO and AFO orders within the ladder approximation. In Sec. \ref{sec:sc} we analyze the pairing instabilities within the same approximation. In Sec. \ref{sec:comp} we compare the instabilities in the particle-hole and particle-particle channels first in the ladder approximation and then by adding RG analysis.  We present our conclusions in Sec. \ref{sec:summary}.

\section{The model}
\label{sec:model}

\begin{figure}[h]
\centering{}\includegraphics[width=0.95\columnwidth]{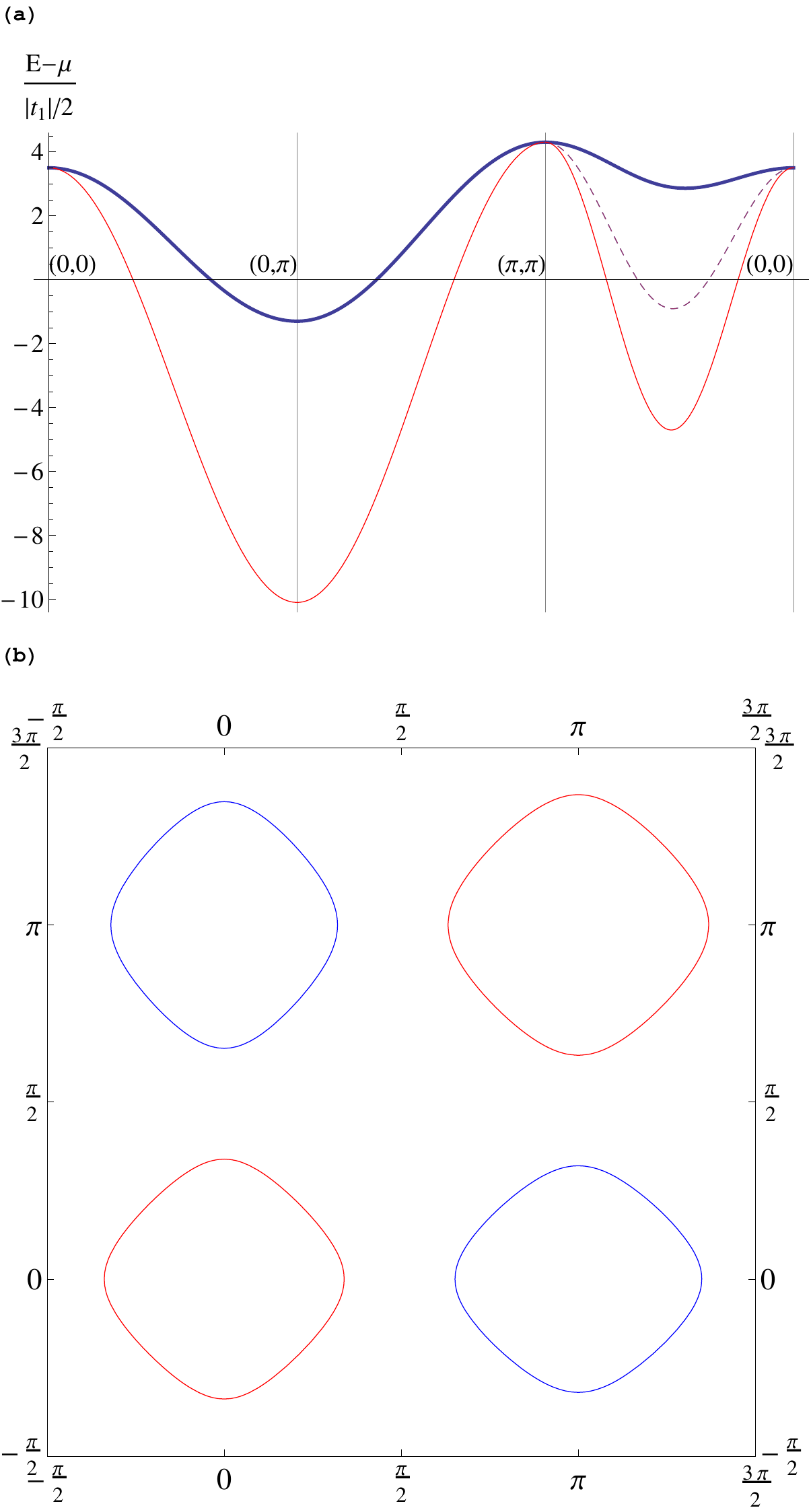} \protect\caption{(a)Dispersion along $(0,0)$-$(0,\pi)$-$(\pi,\pi)$-$(0,0)$ measured from $\mu$. The solid red line shows ${E}_{a}(k)$; the thick blue line shows ${E}_{b}(k)$; the purple dashed line shows $A_{x,k} = A_{y,k}$ from $(\pi,\pi)$ to $(0,0)$. $A_{x,k}$ coincides with ${E}_{a}(k)$ (and $A_{y,k}$ coincides with ${E}_{b}(k)$) along $(0,0)$-$(0,\pi)$ and $(0,\pi)$-$(\pi,\pi)$ directions. We used $t_1=2.0$, $t_2=-2.4$, $t_3=4.8$, $V=3.8$ and $\mu=0.9$.
(b)the Fermi surface of the two-orbital model. The red circles are hole pockets and the blue ones are electron pockets.
\label{fig:dispersion} }
\end{figure}

 We consider the same two-orbital model as in earlier works,  with hoping between $d{xz}$ and $d_{yz}$ orbitals at nearest and next-nearest neighbors.
 The kinetic energy is
 \bea
 {\cal H} &=& \sum_k A_{x,k} d^\dagger_{xz,k} d_{xz,k} + A_{y,k} d^\dagger_{yz,k} d_{yz,k} + \nonumber \\
  &&V_k \left(d^\dagger_{xz,k} d_{yz,k} + d^\dagger_{yz,k} d_{xz,k}\right)
\label{ch_1}
\eea
 where the summation over spin components is assumed and
 \bea
&& A_{x,k} = A_0 + t_1 \cos{k_x} + t_2 \cos{k_y} + t_{3} \cos{k_x} \cos{k_y}, \nonumber \\
&& A_{y,k} = A_0 + t_2 \cos{k_x} + t_1 \cos{k_y} + t_{3} \cos{k_x} \cos{k_y}, \nonumber \\
&& V_k = V \sin{k_x} \sin{k_y}. \nonumber
\eea
  The dispersions $A_{x,k}$ and $A_{y,k}$ along different directions in momentum space are presented in Fig. \ref{fig:dispersion}.
   The two  dispersions are obviously degenerate at $(0,0)$ and at $(\pi,\pi)$. The $V$ term does not remove the degeneracy, but it
     mixes $d_{xz}$ and $d_{yz}$ orbitals away from these points. The kinetic energy in the presence of the $V$ term can be easily diagonalized.
       Near $(0,0)$ and $(\pi,\pi)$ there are two low-energy modes, each with
  mixed $d_{xz}/d_{yz}$ character.  Out of two low-energy modes near $(0,0)$,
   one crosses the chemical potential and creates a hole pocket, while the other remains above the chemical potential. The same holds near $(\pi,\pi)$, where the second hole pocket develops.

 Near ${\bf k} =0$,  $A_{x,k}$, $A_{y,k}$ and $V_k$ are approximated by
 \bea
 && A_{x,k} = \epsilon_0 - a k^2 - c (k^2_x-k^2_y), \nonumber \\
 && A_{y,k} = \epsilon_0 - a k^2 + c (k^2_x-k^2_y), \nonumber \\
 && V_k = V k_x k_y, \nonumber
 \eea
  where
  \beq
  \epsilon_0 = A_0 + t_1 + t_2 + t_{3}, a = \frac{t_1 + t_2 +2 t_{3}}{4}, c = \frac{t_1-t_2}{4}. \nonumber
  \eeq
 The diagonalization near ${\bf k} =0$ yields
 \beq
 {\cal H} = {E}_{a}(k) { a}^\dagger_k { a}_k + {E}_{b}(k) {b}^\dagger_k {b}_k, \nonumber
  \eeq
 and the two dispersions  are
 \beq
 {E}_{a,b}(k) = \epsilon_0 - ak^2 \mp \sqrt{c^2 (k^2_x-k^2_y)^2 + V^2 k^2_xk^2_y}. \nonumber
 \eeq
   To simplify  the analysis we set $V = 2c$, in which case the two dispersions near ${\bf k} =0$ are isotropic: ${\cal H} = {E}_{a}(k) a^\dagger_k a_k + {E}_{b}(k) b^\dagger_k b_k$, where  ${E}_{a,b}(k) = \epsilon_0 - (a \pm |c|) k^2$. The transformation from $d_{xz}/d_{yz}$ orbital operators to $a$ and $b$ band operators is a pure rotation~\cite{Cvetkovic2013}
  \beq
  d_{xz} = a \cos \phi + b \sin \phi, d_{yz} = b \cos \phi - a \sin \phi,
  \label{ch_2}
  \eeq
   where $\phi$ is the angle between ${\bf k}$ and x-axis. Like in earlier works we set $\epsilon_0 >0$, $ a >0$, and $a \sim |c|$.   For these parameters, $E_a$ crosses zero at $k=k_F = (\epsilon_0/(a+ |c|))^{1/2}$, while $E_b$  remains approximately equal to $E_0$ at small $k$.

 A similar analysis for ${\bf k}  \approx  (\pi,\pi)$  yields the similar form of ${\cal H}$ as near ${\bf k} =0$, i.e.,
 \beq
 {\cal H} = {E}_{\tilde a} (k) {\tilde a}^\dagger_k {\tilde a}_k + {E}_{\tilde b} (k) {\tilde b}^\dagger_k {\tilde b}_k, \nonumber
  \eeq
  where ${\bf k}$  is counted from $(\pi,\pi)$, and
  \beq
 {E}_{{\tilde a},{\tilde b}}(k) = {\tilde \epsilon}_0 - (a \pm |c|) k^2 \nonumber
 \eeq
   where $\epsilon_0 = A_0 - t_1 - t_2 + t_{3}$.
  Again, the ${\tilde a}$
 band crosses the chemical potential and forms a hole pocket, while the energy of the ${\tilde b}$ band remains approximately equal to ${\tilde \epsilon}_0$.
   The dispersions near $(\pi,\pi)$  become identical to those near $(0,0)$  when $t_1 +t_2 =0$.
   The transformation from $d_{xz}/d_{yz}$ orbital operators  to ${\tilde a}$ and ${\tilde b}$ band operators is
  \beq
  d_{xz} = {\tilde b} \cos \phi - {\tilde a} \sin \phi, d_{yz} = {\tilde a} \cos \phi + {\tilde b} \sin \phi,
  \label{ch_3}
  \eeq
 where $\phi$ is again the angle between small ${\tilde k}$ and x-axis.  We emphasize that (\ref{ch_3}) is not obtained from (\ref{ch_2}) by rotating $\phi$ by $90^o$, one needs to invoke an additional reflection around, say, $x$ axis.

Near$(0,\pi)$ ($(\pi,0)$), only  $A_{x,k}$ ($A_{y,k}$)  becomes soft, other branch has a larger gap, comparable to the full bandwidth.  The hybridization term $V_k$ vanishes
 at $(\pi,0)$ and $(0,\pi)$, hence low-energy excitations near $(0,\pi)$ ($(\pi,0)$) can be safely approximated  as pure $d_{xz}$ ($d_{yz}$).
  These pure excitations form two electron pockets (see Fig. \ref{fig:dispersion}).
   We label corresponding
  low-energy fermions as $f_{1,k}$ ($f_{2,k}$) with momentum counted from $(0,\pi)$ ($(\pi,0)$).

 We follow Refs.~\cite{ashvin,yamaze} and set the interaction term to be ${\cal H}_{int} = - g \sum_{r,\alpha}(n_{xz,\alpha} (r) - n_{yz\alpha} (r))^2$, where $n_{xz,\alpha} (r) = d^\dagger_{xz,\alpha} (r) d_{xz,\alpha} (r)$ and $n_{yz,\alpha} (r) = d^\dagger_{yz,\alpha} (r) d_{yz,\alpha} (r)$.
  This interaction can be cast into more familiar $U-U'$ Hubbard form with intra-pocket and inter-pocket terms:
 \bea
 {\cal H} &=&   \sum_{r,\alpha,\beta} \frac{U}{2} \left(n_{xz,\alpha} (r)  n_{xz,\beta} (r) + n_{yz,\alpha} (r)  n_{yz,\beta} (r) \right) + \nonumber \\
  && U' n_{xz,\alpha} (r)  n_{yz,\beta} (r)
 \label{ch_6}
  \eea
  with
 $U = -2g$ and $U' = 2g$.   Like in earlier works, we set $g$ to be positive, in which case the interaction favors orbital order with $n_{xz, \alpha} (r) \neq n_{yx,\alpha} (r)$.   The model with a positive $g$ is somewhat artificial as it implies that intra-orbital Hubbard interaction $U$ is attractive, but, like we said, this model allows one to compare QMC results with analytical results at weak and strong coupling.

The interaction (\ref{ch_6})
 is momentum independent in the orbital basis, but acquires the dependence on $\cos{\theta}$ and ${\sin \theta}$ of individual fermions, when re-expressed in the band basis.  Namely, each time $d_{xz}$ or $d_{yz}$ operator is re-expressed in terms of $a,b, {\tilde a}$, or ${\tilde b}$ fermions,  the interaction term acquires the corresponding coherence factor from the transformation from orbital to band basis.

 \section{Instabilities in the particle-hole and particle-particle channels within the itinerant approach}
\label{sec:inst}

 \subsection{Orbital ordering}
\label{sec:oo}

At large $g$, the potential energy well exceeds the kinetic energy. The $-g (n^2_{xz} -n^2_{yz})$ is minimized when all fermions accumulate in one band,  breaking the orbital symmetry.   However, the potential energy is local and it alone does not specify the momentum of the orbital order.  To understand what kind of orbital ordering  develops, one needs to include the leading corrections in $t/g$.  These terms favor a checkerboard, AFO
  order with momentum $(\pi,\pi)$ (Ref.\cite{landau}).  The same AFO order has been found in QMC analysis~\cite{ashvin}.  Like we said, our goal is to understand what kind of orbital order emerges at weaker couplings, when potential energy can be treated as a perturbation and the instability comes from low-energy fermions, located near the Fermi surfaces.

  We compare two types of orbital orders $\Delta (r) = \sum_q \Delta (q) e^{i q r}$: uniform FO order
  $ \Delta (q) =\Delta_{fo} \delta (q)$  and staggered anti-FO order $\Delta (q) = \Delta_{afo} \delta (q -(\pi,\pi))$.
 In terms of low-energy band fermions,
  \begin{widetext}
  \bea
&& \Delta_{fo} =  \sum_k \left[<f^\dagger_{1,k} f_{1,k} -f^\dagger_{2,k} f_{2,k}>\right] \label{ch_4} \\
&& + \sum_k \left[\left(<a^\dagger_k a_k -b^\dagger_k b_k>\right)  \cos{2\theta_k} + \left(<a^\dagger_k b_k + b^\dagger_k a_k>\right)  \sin{2\theta_k}\right] \nonumber \\
&& - \sum_k \left[\left(<{\tilde a}^\dagger_k {\tilde a}_k -{\tilde b}^\dagger_k {\tilde b}_k>\right)  \cos{2\theta_k} + \left(<{\tilde a}^\dagger_k {\tilde b}_k + {\tilde b}^\dagger_k {\tilde a}_k>\right)  \sin{2\theta_k}\right] \nonumber
\eea
and
 \beq
\Delta_{afo}= \sum_k \left[\left(<a^\dagger_k {\tilde b}_k  + {\tilde b}^\dagger_k a_k>\right) - \left(<b^\dagger_k {\tilde a}_k  + {\tilde a}^\dagger_k b_k>\right)\right],
\label{ch_5}
\eeq
\end{widetext}
where the integration over $k$ is confined to the FS and this reduces the integration over position of $k$ on the FS specified by $\theta_k$. In both terms the summation is restricted to small $k$. There is no contribution to $\Delta_{afo}$ from electron pockets  because  out of two fermions from the same orbital, one has high energy.

To understand when (and if) the system develops an instability towards any of these orbital orders, we  add to the Hamiltonian infinitesimally small order parameters
$\Delta^{(0)}_{fo}$ and $\Delta^{(0)}_{afo}$ and
 compute the full susceptibilities. The divergence of a certain susceptibility would signal an instability towards  the corresponding spontaneous order.

\subsubsection{Ferro-orbital order}

 We first do calculations in the ladder approximation and then include RG renormalizations of $U$ and $U'$.
In the ladder approximation (also often called random phase approximation)
one assumes that the dominant contribution to the renormalization of $\Delta^{(0)}_{fo}$ and $\Delta^{(0)}_{afo}$ comes from series of ladder and bubble diagrams with  repeated insertions of the interactions in the same channel, in our case particle-hole channel with momentum transfer either zero or $(\pi,\pi)$, while coupling to other channels (e.h., particle-particle channel) are neglected.  Within this approximation, the
fully renormalized order parameters $\Delta_{fo}$ and $\Delta_{afo}$ are expressed via the bare ones as $\Delta_{fo} = \Delta^{(0)}_{fo}/(1 - I_{fo})$ and $\Delta_{afo} = \Delta^{(0)}_{afo}/(1 - I_{afo})$. The instability in a given channel develops when the corresponding $I=1$.  To obtain when (and if) this condition is satisfied, one can neglect the bare values, find eigenvalues of the self-consistent equations for $\Delta_{fo}$ and $\Delta_{afo}$ and check when the highest eigenvalue reaches one.

\begin{figure}[h]
\centering{}\includegraphics[width=1.0\columnwidth]{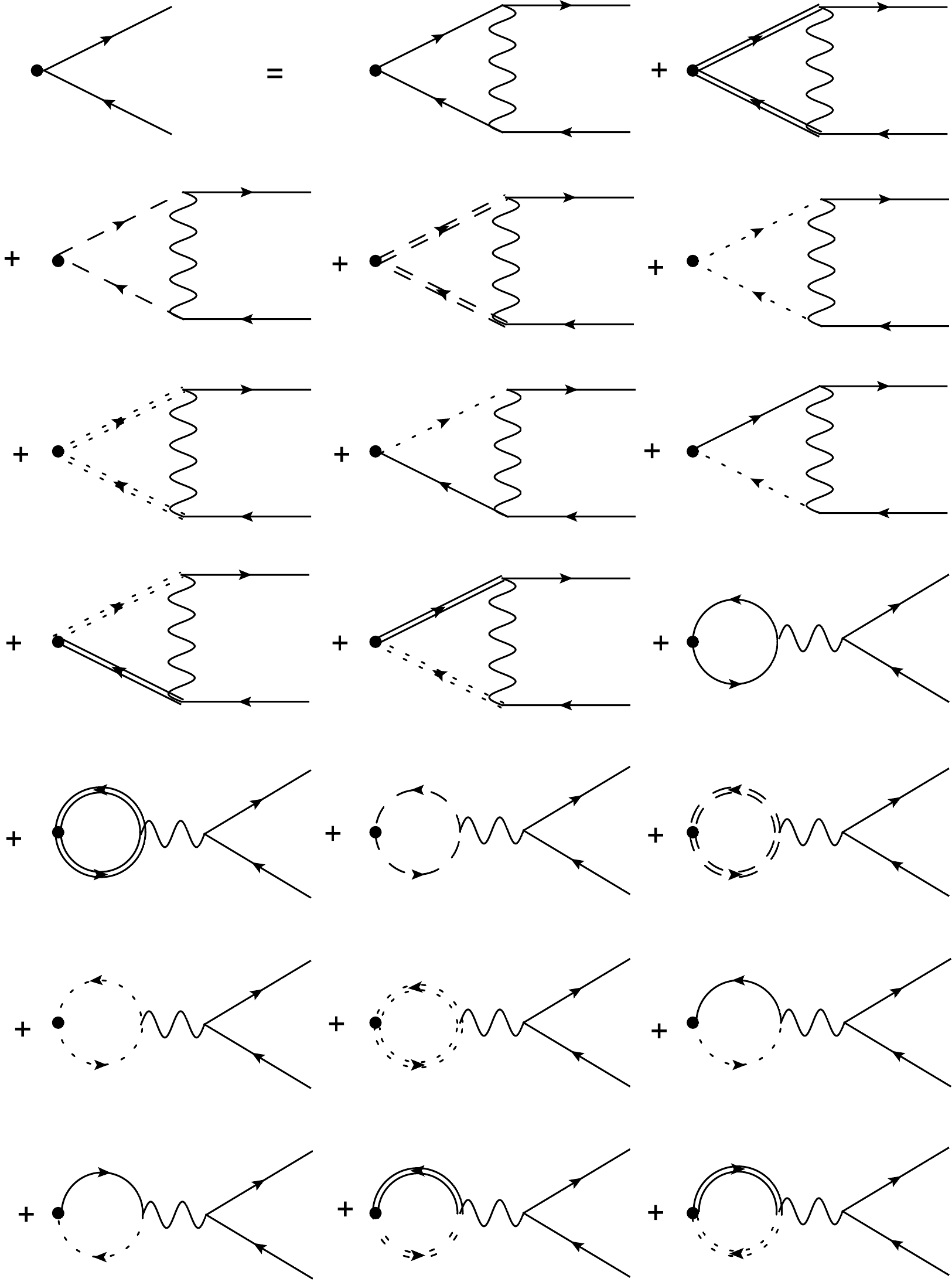} \protect\caption{Diagrams for the renormalization of the components of ferro-orbital order parameter.
Only diagrams for $\Delta^{aa}_{fo}$ are shown; the diagrams for the renormalization of other order parameters $\Delta^{bb}_{fo}$, $\Delta^{ab}_{fo}$, $\Delta^{{\tilde a} {\tilde a}}_{fo}$, $\Delta^{{\tilde b} {\tilde b}}_{fo}$, $\Delta^{{\tilde a} {\tilde b}}_{fo}$, $\Delta^{f_1 f_1}_{fo}$ and $\Delta^{f_2 f_2}_{fo}$ are obtained in a similar way. Solid lines, dotted lines and dashed lines label $a$, $\tilde a$, $f_1$ respectively; double solid lines, dotted lines and dashed lines label $b$, $\tilde b$, $f_2$ respectively.
\label{fig:fig_fo} }
\end{figure}

The set of self-consistent equations for $\Delta_{fo}$ is presented in Fig. \ref{fig:fig_fo}.  Because coherence factors depends separately on $\cos \theta$ and $\sin \theta$, one has to introduce  a more generic $q=0$ order parameter with components
\bea
&&\Delta^{aa}_{fo} \cos^2{\theta},~~{\bar \Delta}^{aa}_{fo} \sin^2{\theta},~~\Delta^{{\tilde a} {\tilde a}}_{fo} \cos^2{\theta},~~{\bar \Delta}^{{\tilde a} {\tilde a}}_{fo} \sin^2{\theta}, \nonumber \\
 &&\Delta^{ab}_{fo} \cos{\theta} \sin{\theta},~~ \Delta^{{\tilde a} {\tilde b}}_{fo} \cos{\theta} \sin{\theta}, \nonumber \\
 &&\Delta^{f_1 f_1}_{fo},~~\Delta^{f_2 f_2}_{fo}, \nonumber \\
  \eea
  where $\Delta^{aa}_{fo} = \sum_k a^\dagger_k a_k$ and so on.  These 8 order parameters are all coupled in the ladder approximation, however the 8 $\times$ 8 secular equation decouples between $s-$wave and two $d-$wave harmonics.  Assume momentarily that $t_1+t_2=0$, i.e., the pockets at $(0,0)$ and $\pi,\pi)$ are identical. Then in the  $d_{x^2-y^2}$ channel
 (the one we need)
\bea
&&\Delta^{aa}_{fo} = -{\bar \Delta}^{aa}_{fo} = -\Delta^{{\tilde a} {\tilde a}}_{fo} = {\bar \Delta|}^{{\tilde a} {\tilde a}}_{fo} = \Delta_{1} \nonumber \\
&& \Delta^{{a} {b}}_{fo} =- \Delta^{{\tilde a} {\tilde b}}_{fo} = \Delta_2, \nonumber \\
&&\Delta^{f_1 f_1}_{fo}  =- \Delta^{f_2 f_2}_{fo} = \Delta_3
\eea
  The three equations on $\Delta_i$, $i=1-3$ are identical up to a factor 2:
  \bea
  &&
  \Delta_1 = \frac{-U + 2 U'}{2} \left[\Delta_1 \Pi_{aa} + 2 \Delta_2 \Pi_{ab} +
  2 \Delta_3 \Pi_{ff}\right] \nonumber \\
 &&\Delta_2 = \frac{-U + 2 U'}{2} \left[\Delta_1 \Pi_{aa} + 2 \Delta_2 \Pi_{ab} +
 2 \Delta_3 \Pi_{ff}\right] \nonumber \\
 &&\Delta_3 = 2 \frac{-U + 2 U'}{2} \left[\Delta_1 \Pi_{aa} + 2 \Delta_2 \Pi_{ab} +
  2 \Delta_3 \Pi_{ff}\right]
 \label{ch_7}
 \eea
 where $\Pi_{ij}$ are polarization operators defined such that $\Pi_{ij} >0$.
  The solution of (\ref{ch_7}) is, obviously, $\Delta_1 = \Delta_2 = \Delta_3/2 = \Delta$.  Substituting this into (\ref{ch_4}) we obtain $\Delta_{fo} = 6 \Delta$.
   The eigenvalue for this solution is
  $\lambda_{fo} = (-U/2+ U') \left(\Pi_{aa} + 2 \Pi_{ff} + 2 \Pi_{ab}\right)$.   For a more generic case when hole pockets are not equivalent, the calculations are a bit more involved, but the result is the expected one:
 \beq
 \lambda_{fo} = \frac{-U + 2 U'}{4}  \left[\left(\Pi_{aa} + \Pi_{{\tilde a}{\tilde a}}\right) + 4 \Pi_{ff} +
  2 \left(\Pi_{ab} + \Pi_{{\tilde a}{\tilde b}}\right)\right]
  \label{ch_8}
  \eeq
 We recall that in our model $U = -2g$ and $U' = 2g$, i.e. $-U+2U' = 6g >0$.  Then, at some critical $g$, the system becomes unstable against FO order.

\subsubsection{Antiferro-orbital order}

\begin{figure}[h]
\centering{}\includegraphics[width=1.0\columnwidth]{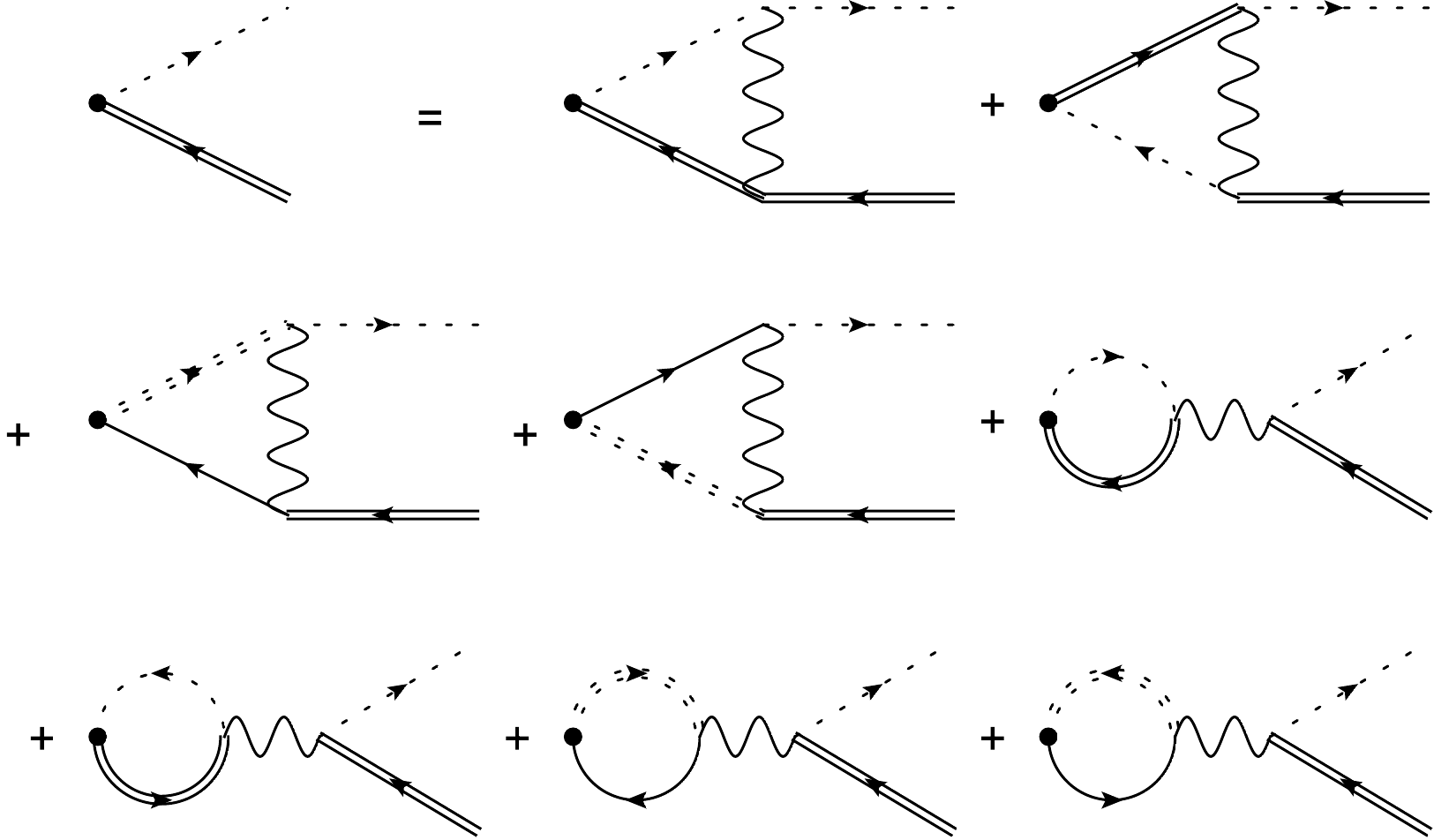} \protect\caption{Diagrams for the renormalization of the components of antiferro-orbital order parameter.
Only diagrams for $\Delta^{{\tilde a} b}_{afo}$ are shown; the diagrams for the renormalization of $\Delta^{a {\tilde b}}_{afo}$ are obtained in a similar way. The notations are the same as in Fig. \ref{fig:fig_fo}.
\label{fig:fig_afo} }
\end{figure}

We now consider AFO order. The set of self-consistent equations for $\Delta_{afo}$ is presented in Fig. \ref{fig:fig_afo}.

  Like before, we have to introduce more general $q=(\pi,\pi)$ order parameters
\bea
&&\Delta^{a{\tilde b}}_{afo} \cos^2{\theta},~~{\bar \Delta}^{a{\tilde b}}_{afo} \sin^2{\theta},~~\Delta^{{\tilde a} b}_{afo} \cos^2{\theta}, ~~
{\bar \Delta}^{{\tilde a} b}_{afo} \sin^2{\theta}, \nonumber \\
&& \Delta^{a {\tilde a}}_{afo} \cos{\theta} \sin{\theta}, ~~\Delta^{f_1 f_2}_{afo}. \nonumber
\eea
  Again, $s-$wave and $d-$wave harmonics decouple. One can straightforwardly verify that
   only the first four parameters contribute to  $d_{x^2-y^2}$ harmonics, and, moreover, in this channel
   \bea
  && \Delta^{a{\tilde b}}_{afo} = {\bar \Delta}^{a{\tilde b}}_{afo} = \Delta_{1a},\nonumber \\
  &&  \Delta^{{\tilde a} b}_{afo} = {\bar \Delta}^{{\tilde a} b}_{afo} = \Delta_{2a}. \nonumber
  \eea
   The coupled equations on
   $\Delta_{1a}$ and $\Delta_{2a}$ are
   \bea
  &&\Delta_{1a} = \left[-\left(U-2U'\right) \Delta_{1a} + \left(U-2U'\right) \Delta_{2a} \right]  \Pi_{a{\tilde b}} \nonumber \\
  &&\Delta_{2a} = - \left[-\left(U-2U'\right) \Delta_{1a} + \left(U-2U'\right) \Delta_{2a} \right]  \Pi_{a{\tilde b}} \nonumber
 \eea
 where we used that $\Pi_{a{\tilde b}}= \Pi_{{\tilde a}b}$. Like before, we defined polarization operator such that $\Pi_{a{\tilde b}} >0$.
  For $U<0$ and $U'>0$ the only positive eigenvalue is
 \beq
 \lambda_{afo} = 2 \left(-U + 2 U'\right) \Pi_{a{\tilde b}}
 \label{ch_10}
 \eeq
 The  corresponding eigenfunction has $\Delta_{1a} = - \Delta_{2a} =\Delta_a$. Substituting into (\ref{ch_5}) we obtain $\Delta_{afo} = 8 \Delta_a$.
 For $U=-2g$, $U' = 2g$, $\lambda_{afo} = 12 g \Pi_{a{\tilde b}}$

 We now compare $\lambda_{fo}$ and $\lambda_{afo}$.   The point for comparison is that for small hole pockets, i.e., for small ratios
  $E_{a,b;0}/W = \epsilon_0/W$ and $E_{{\tilde a},{\tilde b};0}/W = {\tilde \epsilon}/W$,  polarization operators $\Pi_{a b}$, $\Pi_{{\tilde a}{\tilde b}}$, and $\Pi_{a{\tilde b}}$
   are logarithmically enhanced as $\log{W/\epsilon_0} \sim \log{W/{\tilde \epsilon}_0}$, because they are made out of fermions which over wide momentum range have opposite signs of dispersion, i.e., a particle-hole bubble effectively behaves as a particle-particle bubble, up to an overall sign.  As a result, each of these bubbles behaves as $\log{W/\epsilon_0}$.  At the same time, $\Pi_{aa}$ and $\Pi_{ff}$ are ordinary zero-momentum polarization bubbles, and both are of order $1/W$.
     Without $\Pi_{a b}$ and other cross-terms,  $\lambda_{afo}$ would vanish, while $\lambda_{fo}$ would be positive, but of order $g/W$, i.e., there would be no instability at $g \ll W$, where weak coupling approach is justified. Because of cross-terms, the situation is quite different in two aspects. First, the instability occurs at  $g \sim W/\log{\frac{W}{\epsilon_0}} \ll W$, where calculations are under control. Second, the prefactor for the logarithm
      is by a factor of two larger in  $\lambda_{afo}$ than in $\lambda_{fo}$, hence the leading orbital instability is actually towards the AFO order.
      That $\lambda_{afo} > \lambda_{fo}$ is  consistent with QMC results~\cite{ashvin}. QMC calculations show that susceptibility in both channels increases
       as $g$ increases and, at a critical $g_c$,  diverges in the AFO channel, while the susceptibility in the FO channel remains finite  at $g_c$.   The QMC study also found that AFO order develops only at filling when hole pockets are small but finite, and disappears at higher and smaller fillings. This is also consistent with our analysis because at larger electron filling electron pockets grow, and the range where $a$ and ${\tilde b}$
        dispersions have opposite sign shrinks, hence $\Pi_{a{\tilde b}}$ decreases. At large hole doping, $\epsilon_0$ and ${\bar \epsilon}_0$ increase and $\Pi_{a{\tilde b}}$ again decreases, this time because logarithmic enhancement gets weaker.

\subsection{Superconductivity}
\label{sec:sc}

  The same interaction Hamiltonian, Eq. (\ref{ch_6}),
 also gives rise to the SC
instability, and it becomes an issue whether this instability develops before or after AFO order sets in.

The dominant contribution to SC at weak coupling,  when $T_c$ is small enough, comes from states immediately close to the Fermi surface,
i.e., from $a$, ${\tilde a}$, $f_1$ and $f_2$  fermions.  However, when $T_c$ is higher,  one needs to include  the contributions to the pairing from $b$ and ${\tilde b}$ fermions, i.e., particle-particle polarization bubbles $\Pi_{bb}$, $\Pi_{a,b}$  and other terms of the same type.   In the analysis below we keep all contributions in the SC channel.
\begin{figure}[h]
\centering{}\includegraphics[width=1.0\columnwidth]{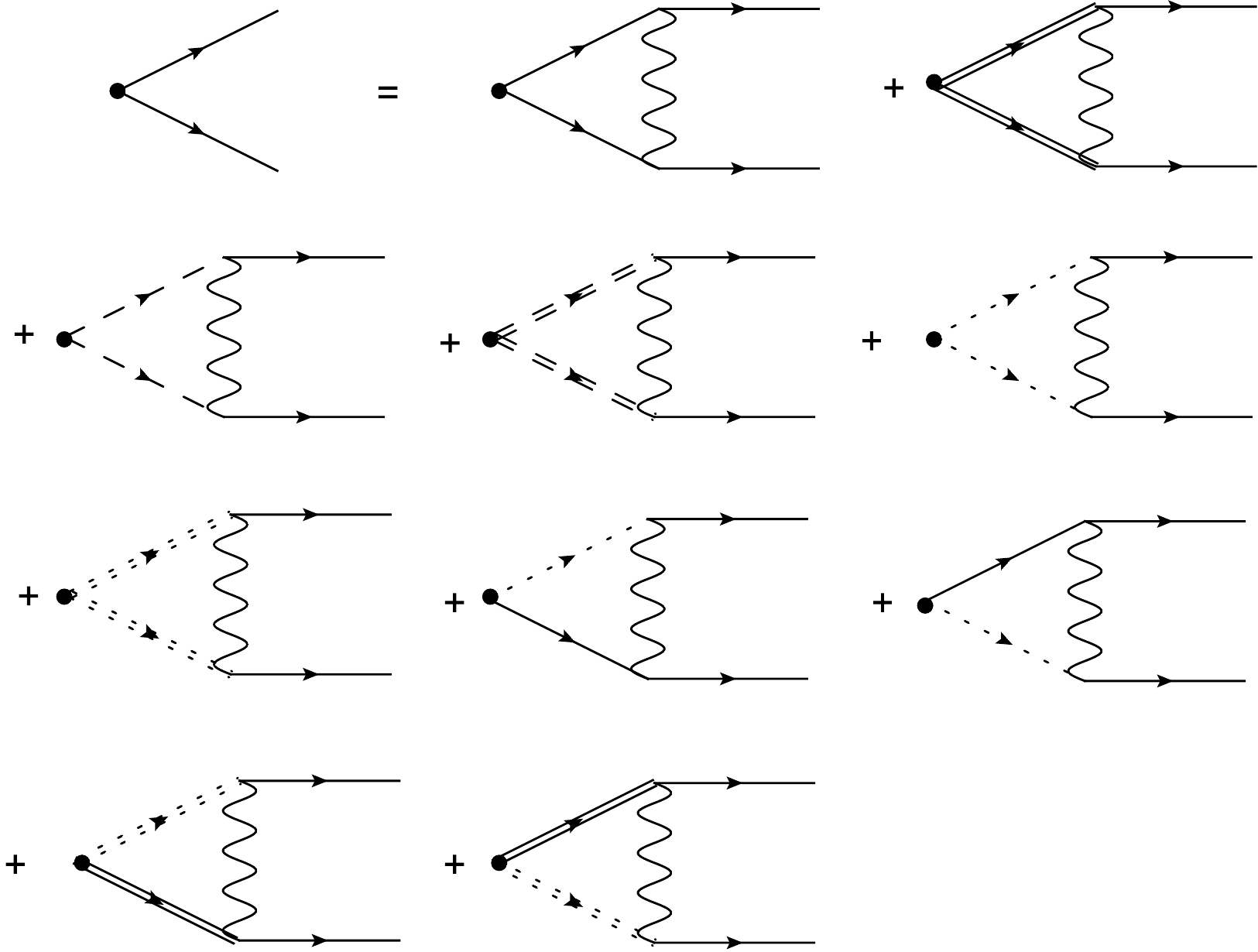} \protect\caption{Diagrams for the renormalization of the components of pairing order parameter.
Only diagrams for $\Delta^{aa}_{SC}$ are shown; the diagrams for the renormalization of other order parameters $\Delta^{bb}_{SC}$, $\Delta^{f_1 f_1}_{SC}$, $\Delta^{f_2 f_2}_{SC}$, $\Delta^{{\tilde a} {\tilde a}}_{SC}$, $\Delta^{{\tilde b} {\tilde b}}_{SC}$, $\Delta^{{\tilde a} {\tilde b}}_{SC}$ and $\Delta^{ab}_{SC}$ are obtained in a similar way. The notations are the same as in Fig. \ref{fig:fig_fo}.
\label{fig:fig_sc} }
\end{figure}
 The calculations are performed in the same way as before(Fig. \ref{fig:fig_sc}), by introducing order parameters
 \beq
 \Delta^{aa}_{SC} \cos^2{\theta},~~\Delta^{aa}_{SC} \sin^2{\theta},~~\Delta^{ab}_{SC} \cos{\theta} \sin{\theta}
 \label{xx}
  \eeq
  and so on,  where $\Delta^{aa}_{SC} = \sum_k a_{k,\alpha} a_{k \beta} (i \sigma^y_{\alpha \beta})$, etc.
   Like before, we derive self-consistent equations on $\Delta^{ij}_{SC}$ in the ladder approximation and obtain eigenvalues. Only the $U$ term contributes to the renormalization of the pairing vertex, $U'$ term doesn't play a role. 
   In total, there are 16 gap components (if we count $\Delta_{ab}$ and $\Delta_{ba}$ as separate variables), i.e., there are 16 coupled equations. 
    By obvious reasons, the equations decouple between 
   $s-$wave and $d-$wave channels. With our choice of variables in Eq. (\ref{xx}), $d-$wave component necessary has $d_{x^2-y^2}$  symmetry. 
    [To analyze the  coupling in $d_{xy}$ channel  one  has to introduce different set of variables like $\Delta^{aa}_{SC} \sin {\theta} \cos {\theta}$, etc.]
    
  The presence of large number of components normally implies that there exist non-zero eigenvalues in different subsets of $s-$wave and $d_{x^2-y^2}$ channels (e.g., $s^{++}$ and $s^{+-}$), and one has to verify which sub-channel wins.  However, we found that in in our case there is no such competition as  eigenvalues are non-zero only in the $s^{++}$ and $d_{x^2-y^2}$ channels.  These two non-zero eigenvalues are  
  \bea
&&\lambda_{s^{++}} = \label{ch_11} \\ &&-2U  \left[\frac{\Pi^{aa}_{pp} + \Pi^{{\tilde a} {\tilde a}}_{pp} + 2\Pi^{ff}_{pp}}{4} + \frac{\Pi^{bb}_{pp} + \Pi^{{\tilde b} {\tilde b}}_{pp}}{4} \right] \nonumber \\
&&\lambda_{d_{x^2-y^2}} = \nonumber \\
&& -U  \left[\frac{\Pi^{aa}_{pp} + \Pi^{{\tilde a} {\tilde a}}_{pp}+4 \Pi^{ff}_{pp}}{4} + \frac{\Pi^{bb}_{pp} + \Pi^{{\tilde b} {\tilde b}}_{pp}}{4}  +\frac{\Pi^{ab}_{pp} +\Pi^{{\tilde a} {\tilde b}}_{pp}}{2} \right] \nonumber
\eea
where $\Pi^{ij}_{pp}$  are particle-particle susceptibilities made out of fermions from band $i$ and $j$ with momenta ${\bf k}$ and $-{\bf k}$, defined such that
$\Pi^{ij}_{pp} >0$.  Note that cross-terms $\Delta^{ab}_{SC} \cos{\theta} \sin{\theta}$ only contribute to $d-$wave channel, and for this channel $\Delta^{ab}_{SC} =- \Delta^{{\tilde a}{\tilde b}}_{SC}$. 
We recall that $U =-2g$ for the model of Eq. (\ref{ch_6}).
 Then $\lambda_{s^{++}}$ and $\lambda_{d_{x^2-y^2}}$ are both  positive, i.e., both channels are attractive.
 For small $g$, the pairing instability occurs at small $T_c$, and the largest contributions to $\lambda$ in both channels comes from $\Pi^{aa}_{pp}$,
 $\Pi^{{\tilde a} {\tilde a}}_{pp}$, and $\Pi^{ff}_{pp}$, which scale as $\frac{1}{W}\log({\frac{W}{T}})$. Other $\Pi^{ij}_{pp}$
  do not diverge at $T=0$, however, because $b$ and ${\tilde b}$ bands are flat over a wide range of momenta, and the band energies in this flat region are of order $\epsilon_0 \ll W$,  $\Pi^{bb}_{pp}$ and $\Pi^{{\tilde b} {\tilde b}}_{pp}$ both scale as $\frac{1}{\epsilon_0} \sim \frac{1}{W} \frac{W}{\epsilon_0} \gg \frac{1}{W}$.
The other two polarization bubbles $\Pi^{ab}_{pp}$ and $\Pi^{{\tilde a} {\tilde b}}_{pp} \sim {\frac{1}{W}}$, i.e., are much smaller.
  Keeping only ${\frac{1}{W}}\log({\frac{W}{T}})$ terms, we find that $\lambda_{s^{++}} > \lambda_{d_{x^2-y^2}}$, i.e., the leading instability in the particle-particle channel is towards $s^{++}$ state.  The $d_{x^2-y^2}$ channel is attractive, but subleading to $s^{++}$.  This result holds when we include $\Pi^{bb}_{pp}$ and $\Pi^{{\tilde b} {\tilde b}}_{pp}$, because the prefactor for $\Pi^{bb}_{pp}$ and $\Pi^{{\tilde b} {\tilde b}}_{pp}$ is larger in the $s^{++}$ channel. Then, the attraction in the $s^{++}$ channel is stronger than in $d_{x^2-y^2}$ channel, no matter what is $T_c$, as long as $T_c \ll W$.

\subsection{Interplay between AFO order and s-wave SC, the role of RG}
 \label{sec:comp}

 Comparing $\lambda_{afo}$ and $\lambda_{s^{++}}$ we find that at the smallest $g$ the system only develops an instability towards $s^{++}$ SC at an exponentially small $T_c$.  If the system is probed by varying $g$ at a given $T \sim \epsilon_0$,  the selection is less obvious because the prefactor $-2U + 4 U'$ in the AFO channel  (Eq. (\ref{ch_10})) is
  larger than $-2U$ for $\lambda_{s^{++}}$ in (\ref{ch_11}), while the combination of the polarization operators is obviously larger in the SC channel.
    This uncertainty goes away once we include the renormalizations neglected in
  the ladder approximation. Specifically, if we apply parquet RG technique for multi-band superconductors~\cite{chubukov}, we find that inter-orbital repulsion $U'>0$  gets renormalized in the particle-particle channel (but not in particle-hole channel) and flows to zero under RG.  This is similar to McMillan-Tolmachev renormalization in conventional phonon superconductor~\cite{mcmillan}.   As the consequence, the prefactor in $\lambda_{afo}$ becomes the same $-2U$ as in  $\lambda_{s^{++}}$. The polarization operators are larger in the SC channel, hence in the ladder approximation, but with running $U$ and $U'$, $s-$wave pairing instability has to develop first, i.e., at a smaller $g$ than  AFO order.
   This is consistent with the results of QMC analysis.
    Another result of RG is that $U$, and, hence, the couplings in both $s^{++}$  and AFO channels get enhanced by the coupling to $(\pi,0)/(0,\pi)$  magnetic fluctuations~\cite{chubukov,frg}. This enhancement is the strongest in the doping range when both hole and electron pockets are small in size. Hence, the instability temperatures are maximized in this region. This again agrees with QMC results.  We caution, however, that using Eqs. (\ref{ch_10}) and (\ref{ch_11}) with the running couplings is an approximation
       not controlled by a small parameter.~\cite{ckf}.

 \section{Summary}
 \label{sec:summary}

  In this paper we analyzed instablities towards orbital order and superconductivity within the two-orbital model for FeSCs, which
   has been recently studied in detail by QMC.  We  used itinerant approach and argued that it is applicable because critical coupling $g$ for orbital
    and  superconducting instabilities is parameterically smaller than the bandwidth.  We found that the leading instability in the orbital channel is towards AFO order with momentum $(\pi,\pi)$, while the one in the pairing channel is towards $s^{++}$ SC, while $d_{x^2-y^2}$ SC is close second.
      We argued that, as $g$ increases at a fixed $T$, the system first develops $s^{++}$ SC order and then, at a larger $g$, develops AFO order.  The latter is confined
       to the range of fillings when hole and electron pockets  are small in size.   The same two orders and the same phase diagram has been recently detected in QMC studies.  We view the agreement with unbiased QMC as the indication that orbital and superconducting orders in FeSCs can be properly accounted for within the itinerant scenario.

 We thank R. Fernandes and A. Vishwanath for useful discussions.  This work was supported
by the Office of Basic Energy Sciences U. S. Department of Energy
under award DE-SC0014402 (AVC).

\end{document}